# Generative artificial intelligence has lowered the barriers to computational social science

Yongjun Zhang
Department of Sociology and Institute for Advanced Computational Science, Stony Brook University, Stony Brook, New York, United States. Direct correspondence to yongjun.zhang@stonybrook.edu

## Abstract
Generative artificial intelligence (AI) has revolutionized the field of computational social science (CSS), unleashing new possibilities for collecting and analyzing multimodal data, especially for scholars who may not have extensive programming expertise. This breakthrough carries profound implications for social scientists. First, generative AI can significantly enhance the productivity of social scientists by automating the generation, annotation, and debugging of code. Second, it empowers researchers to delve into sophisticated data analysis through the innovative use of prompt engineering. Last, the educational sphere of CSS stands to benefit immensely from these tools, given their exceptional ability to annotate and elucidate complex codes for learners, thereby simplifying the learning process and making the technology more accessible.

## Introduction

The rapid advancement of generative artificial intelligence (AI)—exemplified by models such as OpenAI GPT-4 series and Deepseek R1—has significantly transformed the landscape of computational social science (CSS). As an interdisciplinary field, CSS seeks to advance social theory by analyzing large-scale datasets derived from social media, administrative records, and other digitized sources, leveraging a range of computational tools such as natural language processing (NLP) and computer vision techniques (Edelmann et al. 2020; Lazer et al. 2020). One of the complex tasks undertaken by CSS scholars is the collection and distillation of valuable insights from diverse, multimodal data sources—including text, images, audio, and video. For instance, Fan and Zhang (2022) apply structural topic modeling to analyze a large-scale collection of online news articles on COVID19, spanning 65 languages, to examine how foreign media covered China during the pandemic. In a complementary study, Zhang and colleagues analyze millions of tweets by fine-tuning their own large language models to assess public sentiment toward China, revealing the prevalence of Sinophobia in the early stages of the pandemic (Zhang et al. 2023). In the pre-generative AI era, such analyses required scholars to possess advanced programming skills to write custom scripts for web scraping and to fine-tune pre-trained large language and vision models using transfer learning for downstream tasks such as sentiment analysis and image classification.

  Now, the vanguard of generative AI tools like OpenAI GPT series and Meta LlaMa models— empowers scholars to seamlessly convert varied inputs into coherent text outputs. Even more transformative is the capacity for nuanced prompt engineering, enabling a code interpreter that simplifies the generation, annotation, and debugging of code, thereby making complex computational tasks more accessible. The recent development also demonstrates that these tools can be directly used to gather and analyze data with innovative prompt engineering.

In this commentary, I begin by identifying key barriers that have historically limited scholars interested in CSS from acquiring essential technical skills. I then explore how generative AI tools are lowering these barriers by reducing the complexity of coding and streamlining data collection and analysis. Finally, I discuss the epistemic risks associated with the use of generative AI in CSS and offer recommendations for mitigating these challenges.

## What are the main barriers to CSS?

Over the past two decades, CSS has emerged as a growing field in both China and Western societies (Lazer et al. 2020; Wu, Zhang, and Cai 2024). In its early stages, the development of CSS was largely driven by scholars at elite institutions with access to significant resources. While it is unclear who first coined the term "CSS," the field gained formal recognition when David Lazer and his colleagues published a seminal article in Science in 2009 (Lazer et al. 2009). Prior to that, in 2007, a small group of ambitious scientists convened to explore the potential of large-scale data analysis in the social sciences. During that meeting, Harvard political scientist Gary King remarked that the explosion of digital information "will make it possible to learn far more about society and to eventually start solving — actually solving — the major problems that affect the well-being of human populations" (Ledford 2020).

In the late 1980s and early 1990s, the invention of the Internet fundamentally transformed how we live and interact. Since then, life has become increasingly networked and interconnected. As David Lazer and colleagues described in their seminal 2009 Science article,

> "We check out emails regularly, make mobile phone calls from almost any location, swipe transit cards to use public transportation, and make purchases with credit cards. Our movement in public places may be captured by video cameras, and our medical records store as digital files. We may post blog entries accessible to anyone, or maintain friendships through social networks. Each of these transactions leaves digital traces that can be compiled into comprehensive pictures of both individual and group behavior, with the potential to transform our understanding of our lives, organizations, and societies."

This depiction of modern life underscores that we have been living in a digital age for decades. The explosion of big data has provided researchers with powerful new tools to examine human and social behavior through the digital traces we leave behind. These include large-scale archival datasets, administrative records, image and video data, and more. Despite these opportunities, the development of CSS as a formal academic field progressed slowly. Much of the early innovation in CSS occurred within major tech companies such as Google, Facebook, Amazon, and Microsoft, as well as government agencies. It is only in recent years that leading social science disciplines have begun to embrace and publish more research employing computational methods. Over the past decade, CSS has begun to establish a solid foundation in the United States. This growth can be attributed to increased access to large-scale data, more widely available computing resources, advances in analytical methods, and the formation of a robust CSS research community. In the following section, I briefly discuss two major barriers that have historically hindered the adoption of CSS—those related to data collection and analysis.

The explosion of information and communication technologies (ICT) has generated vast amounts of data, but it has also created technical divides. Many traditional social scientists lack the skills to access large-scale structured and unstructured data through methods like web scraping or application programming interfaces (APIs). For example, in my research on social movements in the U.S., a key question is how social movement organizations gain access to policymaking processes (Maher, Seguin, and Zhang 2024; Seguin, Maher, and Zhang 2023). One way to examine this is by analyzing congressional testimonies. To collect these data, we used web scraping techniques on the ProQuest Congressional Hearing Database, available through the University of Arizona. We extracted metadata from every listed hearing between 1946 and 2016, including information on committees, hearing dates, and details about nearly 880,000 testimonies across over 91,000 hearings—such as witness names, affiliations, and titles. In the pre-generative AI era, scraping data from dynamic websites required advanced technical skills, including writing Python or R code with Selenium to automate web browsers. This posed a significant barrier for scholars without computer science training. Even API access—an ostensibly more accessible method for collecting data from public websites or social media platforms—still requires considerable programming expertise. For instance, our team is developing a data infrastructure to track longitudinal YouTube activity related to U.S. presidential elections. Using the YouTube Data API v3, we query the platform daily to collect metadata on millions of videos, including views, likes, dislikes, comments, thumbnails, and channel information. Executing this project requires sophisticated Python scripts to query the API, filter relevant content, manage data storage on remote servers, and handle unexpected errors—tasks that are prohibitively complex for many social scientists without technical training.

The growing accessibility of multimodal data presents exciting opportunities for social scientists, but effectively leveraging these resources often requires advanced computational methods such as NLP and computer vision. For example, urban scholars increasingly use satellite imagery to estimate poverty and wealth, especially in regions where traditional economic data are sparse. In many developing countries, the absence of nationally representative surveys makes it difficult for organizations like the United Nations to target poverty alleviation efforts. To address this gap, researchers have used daytime satellite imagery to estimate economic well-being with promising accuracy (Burke et al. 2021; Jean et al. 2016). Traditionally, analyzing large-scale imagery data required coding expertise in frameworks like PyTorch or TensorFlow to train convolutional neural networks (CNNs). Advances in deep learning, particularly the development of architectures such as GoogLeNet, VGG, ResNet, and Inception, have made it possible to extract meaningful features from images. More recently, transformer-based models like the Swin Transformer have achieved state-of-the-art performance in computer vision tasks (Liu et al. 2021). However, social scientists rarely train large neural networks from scratch due to limited access to labeled data. Transfer learning offers a practical solution: pretrained models can be repurposed to extract low-level features from new images (Wankmüller 2022). Jean et al. (2016) applied this strategy using data from five African countries—Nigeria, Tanzania, Uganda, Malawi, and Rwanda—to predict local economic outcomes. Starting with a CNN pretrained on ImageNet, they fine-tuned it to predict nighttime light intensity from daytime satellite imagery. Nightlights serve as a globally available, albeit imperfect, proxy for economic activity. In this intermediate step, the model learned to reduce high-dimensional input images to compact feature vectors predictive of nightlight intensity. These feature vectors were then used, alongside cluster-level survey data, to train ridge regression models estimating economic indicators such as expenditures or assets. Although transfer learning spares scholars from training deep learning models from scratch, fine-tuning these models still requires writing PyTorch or TensorFlow code—a significant barrier in the pre-generative AI era.

In summary, the major hurdle for social scientists entering computational research has been the need to write complex code for tasks such as collecting large-scale online data and extracting key insights from multimodal data such as text and images. Before the rise of generative AI, researchers commonly relied on tools like Python Selenium to scrape dynamic websites and fine-tuned pretrained models—such as BERT for text and ResNet or Swin Transformer for images—for classification tasks. These processes required proficiency in frameworks like PyTorch or TensorFlow, creating a steep learning curve for scholars new to CSS. Next, I will illustrate how generative AI dismantle these barriers by reducing the complexity of coding and streamlining data collection and analysis.

**Coding is no longer a big challenge to CSS scholars.**
In the era of generative AI, marked by the advent of OpenAI's GPT models, I argue that coding should no longer be the primary barrier for scholars interested in CSS. Generative AI tools can now assist with generating, annotating, and debugging code for a wide range of data collection and analysis tasks. This is not to suggest that scholars no longer need to learn CSS methods, but rather that the technical entry point has shifted—enabling more researchers to engage with computational approaches without being overwhelmed by the coding itself.

*Generating Code*
A pivotal advancement in the realm of generative AI has been its capacity to craft functional code for end-users. Tools such as GitHub Copilot, Code LLaMa, and the ChatGPT code interpreter have become staples for scholars and engineers, facilitating daily code generation. This proves especially advantageous for social scientists whose expertise in programming may be nascent. I will demonstrate this through OpenAI's ChatGPT (GPT-4), highlighting its code interpreter's ability to generate scripts that fine-tune a Swin Transformer model for image classification in the context of studying social protests.

To illustrate, consider my project where I gathered a dataset of over 2 million images from around the globe, potentially depicting social protests. These images reside on my network attached storage, organized within a directory named *images by country*, further segmented into subdirectories labeled by country codes. Additionally, I have leveraged the UCLA protest image dataset, which comprises 40,764 images — 11,659 of which are protest-related — each tagged with various visual attributes and sentiments (Won et al. 2017). It's noteworthy that while the images in the training dataset are sorted into train and test folders, the corresponding labels are housed in distinct train-annotation and test-annotation TSV files. These files contain variables including fname and protest labels, where fname matches the image filenames in the train and test directories. The core objectives are two-fold: first, to script Python code that loads the UCLA protest image dataset to refine a pre-trained Swin Transformer v2 model for protest image identification, and second, to apply the refined model to classify the 2 million image dataset.

The following section outlines the initial prompting process used to complete the task. A close examination of the generated code (not shown in the main text) indicates that ChatGPT "understood" the task in a clear and structured manner. While not without issues, the code produced was generally of high quality.

Prompt: Write pytorch code to fine tune a swin transformer model v2. Training images are stored in train and test folders while labels are stored in two separate tsv files train_annot.txt and test_annot.txt. Note that these txt files contain variables like fname and protest. fname is file names corresponding to images in train and test folders. You need to write python code to load image and label data and then finetune, evaluate, and save swin transformer model. Finally, you need to use the saved model to infer images stored in images_by_country with subfolders named by countries. Make sure you write python codes to check whether these images files are valid.

ChatGPT: Below is a Python code to fine-tune a Swin Transformer model using PyTorch. This code assumes you have the necessary environment and the Swin Transformer pre-trained model installed.
The steps are as follows:
Read the annotation files and prepare the datasets.
Define a custom dataset class to load images and labels.
Define the model, loss function, and optimizer.
Train the model.
Evaluate the model on the test dataset.
Save the fine-tuned model.
Infer on new images from images_by_country and check for valid image files.
First, make sure you have installed the required libraries:
pip install torch torchvision timm pandas pillow
Here's the complete Python code…
(See Appendix, not shown due to space limits)

### *Annotating Code*

Generative AI tools can also assist social scientists by annotating code. The initial output may include complex or non-operational snippets, which can be especially challenging for those new to coding. Understanding function logic and parameter settings can be daunting. To bridge this gap, one can prompt a code interpreter like ChatGPT to break down and explain the code line by line, enabling a clearer grasp of technical details. Below is an example of how ChatGPT can be engaged for detailed code annotation.

"I've received this block of code from the initial output, but as someone new to programming, I'm struggling to grasp its structure and logic. Could you walk me through this code, providing annotations for each segment and explaining the purpose of the different parameters? This way, I can better understand what each part of the code is intended to do and how I might adjust it to fit my specific needs."

By posing the request in this manner, you are likely to receive a breakdown that not only sheds light on the code's current functionality but also imparts knowledge that can empower you to modify and adapt the code independently in the future. Here is the detailed example.

Prompt: I don't understand the train function in the above generated code. Please annotate the code line by line and explain it to me in detail. Make sure you explain the parameters in those functions.

ChatGPT: Certainly! Here is a detailed annotation of the train function, breaking down each line and explaining what it does:

```
def train(dataloader, model, loss_fn, optimizer):
    """
    The train function is responsible for one epoch of training the model.
    Parameters:
    - dataloader: An instance of DataLoader which provides batches of the dataset.
    - model: The neural network model that is being trained.
    - loss_fn: The loss function that is used to evaluate the performance of the model.
    - optimizer: The optimization algorithm used to update the weights of the model.
    """
    # Size of the dataset being used for training.
    size = len(dataloader.dataset)
    # Set the model into training mode which enables the training specific operations like dropout.
    model.train()

    # Loop through each batch in the DataLoader.
    for batch, (X, y) in enumerate(dataloader):
        # X holds the batch of images, y holds the corresponding labels.
        # Move the input and target batches to the same device as the model (GPU or CPU).
        X, y = X.to(device), y.to(device)
        # Perform a forward pass: compute the predicted outputs by passing inputs to the model.
        pred = model(X)
        # Compute the loss by comparing the predictions to the actual labels.
        # The unsqueeze operation is used to add a dimension to y so it matches the shape of pred.
        loss = loss_fn(pred, y.unsqueeze(1).type_as(pred))
        # Before the backward pass, use the optimizer to zero out any gradients from the previous step
        # to prevent accumulation from previous forward passes.
        optimizer.zero_grad()
        # Perform a backward pass: compute the gradient of the loss with respect to model parameters.
        loss.backward()
```

```
        # After calculating the gradients, we update the parameters with the optimizer step.
        optimizer.step()
        # Print out the loss every 100 batches to check the training progress.
        # This is not strictly necessary but helpful to know that something is happening.
        if batch % 100 == 0:
            loss, current = loss.item(), batch * len(X)
            print(f"loss: {loss:>7f}  [{current:>5d}/{size:>5d}]")
```

(The complete output was not shown here due to space limits)

### *Debugging Code*

When the initial code output from ChatGPT doesn't fully align with your requirements, it's prudent to iterate using more specific and targeted prompts. This back-and-forth process is essential for refining the code to better suit your goals and constraints. Through each iteration, you can gradually shape the output to be more efficient, relevant, and aligned with best practices in your domain.

For example, the original code may omit the use of the transformers library from HuggingFace, which is a widely adopted tool in NLP tasks. Additionally, the initial code might implement training loops manually when the same task could be simplified by leveraging the built-in Trainer class from the transformers library. Recognizing these opportunities for optimization requires some domain knowledge, but once identified, ChatGPT can be prompted to incorporate these changes with ease.

Here's an example of a refined prompts used to guide further code development:

Prompt: Revise the above code using transformer library from huggingface and make sure you use trainer function instead of pytorch code.

```
ChatGPT: Here is a revised version of the script:
from transformers import SwinForImageClassification, SwinProcessor, Trainer, TrainingArguments
from datasets import load_metric
from torch.utils.data import DataLoader
from PIL import Image
import pandas as pd
import torch
import os
from concurrent.futures import ThreadPoolExecutor
# Custom dataset class
(ProtestDataset class object was omitted due to space limits.)
# Load the processor and model
processor = SwinProcessor.from_pretrained('microsoft/swin-base-patch4-window7-224')
model = SwinForImageClassification.from_pretrained('microsoft/swin-base-patch4-window7-224', num_labels=2)
# Prepare datasets
train_dataset = ProtestDataset(img_dir='train', annotations_file='train_annot.txt', processor=processor)
eval_dataset = ProtestDataset(img_dir='test', annotations_file='test_annot.txt', processor=processor)
# Define the training arguments
training_args = TrainingArguments(
    output_dir='./results',
    evaluation_strategy="epoch",
    learning_rate=2e-5,
    per_device_train_batch_size=8,
```

```
    per_device_eval_batch_size=16,
    num_train_epochs=3,
    weight_decay=0.01,
)
# Initialize the Trainer
trainer = Trainer(
    model=model,
    args=training_args,
    train_dataset=train_dataset,
    eval_dataset=eval_dataset,
    compute_metrics=load_metric("accuracy"),
)
# Train and evaluate
trainer.train()
trainer.evaluate()
# Save the fine-tuned model
trainer.save_model("swin_transformer_finetuned")
```

I ran the above codes using a workstation with two GPUs and 128 core CPUs. It returns the following error.

Code bug: ImportError: cannot import name 'SwinProcessor' from 'transformers'

I used the following prompt to ask ChatGPT to fix the code for me.

Prompt: After running your code, I got the following error: ImportError: cannot import name 'SwinProcessor' from 'transformers'. Fix it.

ChatGPT: I apologize for the confusion earlier. As of my last update, the transformers library does not include a SwinProcessor. Instead, you would use a feature extractor designed for image models. Let's correct the code by using SwinForImageClassification with an appropriate feature extractor.

```
# Load the feature extractor
feature_extractor = AutoFeatureExtractor.from_pretrained('microsoft/swin-base-patch4-window7-224')
```

After several rounds of iterative debugging, the refined Swin Transformer model impressively attained an accuracy rate of 94% in classifying whether an image is protest-related. The polished codes including the training and inference of these images are detailed in the Appendix.

**Generative AI can help social scientists gather and process multimodal data directly.**

The preceding example highlights the potential of generative AI tools to streamline the coding process, particularly in tasks such as fine-tuning advanced pre-trained vision models for image classification. While I argue that coding is no longer a prohibitive barrier to conducting computational research, it remains essential for social scientists to develop a foundational understanding of these emerging technologies. Beyond facilitating code generation, generative AI also offers the ability to directly collect and analyze multimodal data through effective prompt engineering. A prompt refers to a sequence of text input that instructs large language models (LLMs) to perform specific tasks, enabling researchers to bypass traditional model fine-tuning for many downstream applications. Prompt engineering, in turn, involves crafting and refining these prompts to guide the model toward producing accurate and relevant outputs.

*Data retrieval*
In addition to using scripts generated by GAI tools to scrape online data, scholars can leverage these tools to search the web directly. Some versions of OpenAI's models (e.g., gpt-4o-search-preview), for instance, are equipped with browsing capabilities, allowing researchers to access up-to-date information from the internet before generating a response. This functionality is especially valuable for retrieving the latest data, monitoring real-time trends, or verifying current facts—tasks that are critical in fast-moving research areas. Of course, scholars need to pay attention to the possibility of model hallucination.

Below is an example I use GPT-4o search function to retrieve data on recent U.S. protests.

Prompt: You are an expert social movement scholar and trying to search the web to identify all protest events occurring yesterday in the United States. Please retrieve these events by saving urls, content, and summarizing key event information in these websites. You need to organize the output into a structured table.

ChatGPT: Here is a structured summary of notable protest events that occurred on April 15, 2025, in the United States.

| Location | Event Description | Issues Raised | Source(s) |
|---|---|---|---|
| Acworth, Georgia | At a town hall hosted by Rep. Marjorie Taylor Greene, police removed at least nine protesters; three were arrested, and two were tased during confrontations. Greene defended the removals, asserting that the protesters were out of line. | Opposition to Greene's support for President Trump's policies and recent controversial statements. | People, Reuters, ABC News |
| New York City, New York | Hundreds of Jewish protesters held a "Liberation Seder" outside ICE headquarters, demanding the release of foreign students detained by ICE. | Concerns over the detention of international students and broader immigration policies. | Democracy Now |
| **State College, Pennsylvania** | Approximately 50 Penn State University students protested at the Allen Street Gates, supporting international students and immigrants. The demonstration called on the university to protect students from deportation. | Opposition to ICE activities and support for immigrant rights. | Wikipedia |
| **Bloomington, Indiana** | Around 100 protesters gathered outside the Monroe County Convention Center during an event featuring Senator Todd Young. Demonstrators expressed opposition to the Trump administration's policies. | Concerns over potential cuts to social safety nets and opposition to Trump's agenda. | Indiana Daily Student |
| **Chicago, Illinois** | A "Tax Day March and Rally" was held, starting at 630 W Harrison and proceeding to the South Loop Tesla Dealership. Participants advocated for taxing the wealthy and investing in community services. | Economic inequality, taxation policies, and opposition to budget cuts affecting healthcare and education. | Titan Security |

When scholars use generative AI's web search and text summarization features, the results may not always be comprehensive. However, these tools offer a valuable starting point—especially for those who are not experts in writing sophisticated Python code.

### *Data annotation*

The rapid advancement of generative AI has also enabled these models to process multimodal data directly—including text, images, audio, and video—with increasing sophistication. A case in point is the enhanced GPT-4 model with vision capabilities. Consider an image from the Women's March in Washington, sourced from National Public Radio (NPR), which captures a large crowd unified in their call for women's rights. By uploading this image to GPT-4 and providing targeted prompts, I was able to extract key information efficiently. This example illustrates not only the model's capacity to interpret complex visual data, but also its potential as a powerful tool for social scientists to generate insights without requiring deep technical expertise.

Prompt 1: Describe the image using concise sentences. Then tell me if it is protest-related and show me your reasoning.

ChatGPT: The image shows a group of people actively participating in a march. They are holding a large banner that reads "END WOMEN'S WAGE DISC." and "2019 WOMEN'S MARCH ON WASHINGTON", indicating the event is the 2019 Women's March in Washington D.C. The participants are carrying signs with various slogans and a few are wearing pink hats, which are symbols associated with the Women's March movement. One person in the foreground is using a megaphone, suggesting speeches or chants are part of this event. This is definitely protest-related, as the Women's March is a well-known event advocating for women's rights and gender equality. The presence of signs, a large banner with a clear message regarding women's wage disparity, and the collective action of marching are consistent with protest activities.

Prompt 2: Summarize the image and tell me the sentiment, content, and crowd size.

ChatGPT: The image captures a moment from the 2019 Women's March in Washington D.C., with participants holding a banner advocating against women's wage discrimination. The sentiment appears to be one of solidarity and activism, aimed at promoting gender equality and social justice. The signs and symbols, including pink hats and raised fists, communicate a demand for change and unity among the marchers. Content-wise, the photo includes a diversity of participants of various ages and ethnicities, and some of the visible signs carry messages of empowerment and resistance. The crowd size in the immediate vicinity of the banner is substantial, with dozens of people visible. However, the photograph does not show the entire scope of the march, so the total number of participants is likely much larger, potentially in the thousands, which is typical for such events.

### *Data analysis*

For scholars aiming to process a multitude of images, the integration of OpenAI's GPT-4 with its vision API emerges as a formidable tool. While access to this service is contingent upon a subscription, the investment unlocks GPT-4's sophisticated data analytics capabilities. Social scientists can harness these features for elementary data analysis tasks, thereby enhancing their research productivity significantly.

Take, for instance, my utilization of GPT-4 to dissect the Dynamics of Collective Action dataset, culminating in a detailed map that delineates the geographical disparities in protest occurrences throughout the United States. For an in-depth understanding of the DoCA dataset, visit the website hosted by Stanford University. Upon uploading a segment of this dataset to GPT-4 and prompting it appropriately, I could extract intricate data visualizations. Notably, within the realm of data governance, users have the autonomy to restrict GPT-4 from employing their data in model training, ensuring control and confidentiality.

Prompt: Describe the CSV file and summarize the data pattern.

ChatGPT: I'll first need to load and examine the contents of the CSV file to provide you with a description and summarize the data pattern. Let me do that for you.

```
import pandas as pd
# Load the CSV file into a DataFrame
file_path = '/mnt/data/data.csv'
data_df = pd.read_csv(file_path)
# Display the first few rows of the DataFrame to understand its structure
data_df.head()
```
The CSV file contains a dataset with the following columns:
rptyy: Appears to represent a year, with floating-point numbers (likely intended to be integers). eventid: An identifier for an event, also numeric.
state1: A text field containing abbreviations of US states. city1: A text field containing city names.
The data pattern suggests that this file is a record of events, with each row representing a specific event identified by eventid, occurring in a particular city1 and state1 during a given rptyy (year).
To summarize the data pattern further, I can provide statistics like the number of unique years, event counts, and the frequency of events by state and city. Would you like me to do that?

Prompt: Yes. Please provide the frequency of events by state.

ChatGPT: The frequency of events by state is as follows:
New York (NY): 8,613 events
Washington D.C. (DC): 1,902 events
California (CA): 1,398 events New Jersey (NJ): 1,371 events
Massachusetts (MA): 881 events
Illinois (IL): 772 events

**The epistemic risks of using generative AI in CSS**

Undoubtedly, generative AI has lowered the entry barriers to CSS by reducing coding complexity, but its integration into social science research introduces significant epistemic risks—particularly when large language models (LLMs) are adopted uncritically (Bail 2024). As Bail warns, the unreflective rush to embrace generative AI may lead to the production of "junk science." While some studies have shown that tools like ChatGPT can outperform human crowd workers on specific tasks such as text summarization (Gilardi, Alizadeh, and Kubli 2023), systematic evaluations reveal important limitations. For instance, Hao and Zhang (2025) demonstrate that outputs generated by LLMs are highly sensitive to prompt engineering, raising concerns about reproducibility and validity in data annotation workflows.

To address these challenges, computer scientists have proposed a variety of prompting strategies—such as in-context learning, chain-of-thought, tree-of-thought, and role-play prompting—to improve model performance (Wang et al. 2023; Yao et al. 2023). Yet, within the CSS community, there remains little consensus on best practices for integrating these techniques into research design. While generative AI tools can aid in coding for data collection and analysis, their reliability and replicability remain uncertain due to phenomena like model hallucination. Notably, research has shown that identical prompts with the same parameter settings can yield divergent outputs, further exacerbating the replication crisis already present in the social sciences. Moreover, the increasing reliance on proprietary LLMs developed by leading AI firms—such as OpenAI, Google, and Anthropic—poses additional challenges. These models are subject to "model drift" over time, and their architectures are often opaque, hindering efforts to reproduce results (Ollion et al. 2024). The potential use of sensitive social science data to train such models also raises ethical concerns about data privacy and consent. Taken together, these issues underscore the urgent need for critical scrutiny and methodological rigor in the use of generative AI within social science research.

Although generative AI has lowered the barriers to CSS, it is essential that scholars engage with these tools critically and with caution. A basic conceptual understanding of the underlying computational methods—along with their strengths and limitations—is still necessary. Without such understanding, there is a heightened risk of misapplication, leading to the classic problem of "garbage in, garbage out," where flawed inputs or inappropriate use of tools yield misleading or invalid results. Here, I present several recommendations based on Hao and Zhang (2025) for effectively using generative AI in direct data collection and analysis. First, when employing generative AI's web search capabilities for data retrieval, it is crucial to ensure that the models are prompted to save the URLs and source information. This step allows researchers to validate the retrieved content and maintain transparency in their data collection process. Second, when we are directly using generative AI tools for data annotation, we need to build a development dataset to systemically test potential prompts. Scholars need to consider different parameter settings and prompting strategies for identifying the optimal prompt based on your development dataset. Third, scholars should decompose complex tasks into simple ones to increase the performance, as generative AI tools tend to perform well in simple tasks such as binary classification. Fourth, scholars can explicitly ask generative AI models to output intermediate reasoning steps, which can potentially increase the transparency of the decision-making processes in model reasoning. Fifth, scholars should prioritize open-sourced models such as Meta Llama series and Deepseek R1 models to protect data privacy issues and increase replicability. This is especially crucial if scholars are concerned about uploading their sensitive data to external platforms. Finally, cross validation is essential in any CSS research as we cannot trust generative AI results blindly. Scholars always need to develop an evaluation dataset with human coding to evaluate the model performance and report those key metrics for the community. Together, these recommendations underscore the importance of methodological rigor and ethical awareness as generative AI becomes more embedded in the social science research pipeline.

**The future of CSS in the era of generative AI**
In this commentary, I have attempted to illustrate how the advent of generative AI has democratized computational methodologies, enabling social scientists, even those with minimal coding proficiency, to harness the power of cutting-edge computational tools. The crux of this technological leap is that coding, once a formidable barrier to entry in the digital humanities and CSS, is being redefined as a facilitator rather than a gatekeeper.

 Generative AI is a game-changer for social sciences, as posited by Bail (2024) and Davidson (2024). Its primary benefit lies in alleviating the coding burden, thereby unlocking higher productivity and efficiency among researchers. This liberation from the intricacies of code generation allows social scientists to invest more deeply in the conceptualization of their research, potentially fostering the advancement of social theories. Moreover, the transformative influence of generative AI has reshaped the research paradigm, making computational tools accessible to a broader spectrum of scholars, thereby catalyzing discoveries in social science. Significantly, these tools have also proven to be exceptional assets in pedagogy, offering substantial capabilities to annotate and clarify coding concepts for students in CSS disciplines.

 However, this progress is not without its caveats and conditions. Despite the argument that coding should not pose a substantial hurdle, scholars must still possess fundamental skills to assess both code and the quality of its outputs. For example, when generating code for models like the Swin Transformer, a foundational understanding of deep learning principles is essential. Thankfully, a basic course in CSS is typically sufficient to surmount this learning curve. A more pressing concern is the evaluation of outputs, particularly when employing models such as GPT-4 for vision and advanced data analysis tasks. For those dealing with extensive image datasets, establishing protocols to systematically verify the accuracy of the results is imperative and an area ripe for scholarly inquiry.

While some of the tools referenced here are proprietary, budgetary constraints need not stifle scholarly ambition. Researchers with a modicum of coding knowledge can turn to open source alternatives like LLaMa series, which do require some computing resources to execute code locally. For those without access to significant computational power, free resources such as Google Colab present a viable solution, thereby continuing to lower the barriers and broaden the horizons for research in the social sciences.

## Acknowledgments

I acknowledge the support from the Institute for Advanced Computational Science at Stony Brook University for access to high performance computing systems and OpenAI APIs and ChatGPT.

## Appendix

After several rounds of iteration, I successfully developed a set of functional scripts to finetune a Swin Transformer model and apply it for inference on a new dataset. The code is organized into two separate scripts for clarity and modularity.

```python
from urllib.request import urlopen
import os
import pandas as pd
import torch
from torch.utils.data import Dataset, DataLoader
from torchvision.io import read_image
from transformers import Swinv2ForImageClassification, Swinv2Config, AutoImageProcessor, Trainer, TrainingArguments
from transformers import DataCollatorWithPadding
from torch.utils.data.dataloader import default_collate
import numpy as np  # Make sure numpy is imported

# Custom dataset class
class CustomImageDataset(Dataset):
    def __init__(self, annotations_file, img_dir, transform=None, target_transform=None):
        self.img_labels = pd.read_csv(annotations_file, sep="\t")[['fname', 'protest']]
        self.img_dir = img_dir
        self.transform = transform
        self.target_transform = target_transform

    def __len__(self):
        return len(self.img_labels)

    def __getitem__(self, idx):
        img_path = os.path.join(self.img_dir, self.img_labels.iloc[idx, 0])
        image = read_image(img_path).float()  # Convert to float for ensuring float32 precision

        # Ensure image is in CHW format expected by PyTorch models
        if image.ndim != 3:
            raise ValueError(f"Image at {img_path} does not have three dimensions. Found {image.ndim} dimensions.")
```

```python
        label = int(self.img_labels.iloc[idx, 1])

        # Apply the image processor. It handles necessary image transformations
        processed_images = self.transform(image, return_tensors="pt")

        # SwinV2 model expects keys 'pixel_values' for images. Also, we don't need to unsqueeze here as 'return_tensors' takes care of it
        return {
            'pixel_values': processed_images['pixel_values'].squeeze(0),  # remove the added batch dimension
            'labels': torch.tensor(label)  # Ensure label is a tensor
        }

def custom_data_collator(features):
    # 'features' is a list of dicts with the outputs of your dataset's __getitem__ method
    # You need to aggregate the different elements of these dicts here, for instance by creating a new dict with aggregated tensors

    batch = {}

    # Aggregate the 'pixel_values' and 'labels' from each feature in the batch
    batch['pixel_values'] = torch.stack([f['pixel_values'] for f in features])
    batch['labels'] = torch.tensor([f['labels'] for f in features])

    return batch

# Load the Swin Transformer V2 model from HuggingFace
model_name = "microsoft/swinv2-base-patch4-window8-256"
config = Swinv2Config.from_pretrained(model_name)
config.num_labels = 2  # Set the number of output classes
model = Swinv2ForImageClassification.from_pretrained(model_name, config=config,ignore_mismatched_sizes=True)

# Initialize the AutoImageProcessor
processor = AutoImageProcessor.from_pretrained(model_name)

# Create dataset instances
train_dataset = CustomImageDataset(annotations_file='./annot_train.txt', img_dir='./img/train', transform=processor)
test_dataset = CustomImageDataset(annotations_file='./annot_test.txt', img_dir='./img/test', transform=processor)

# Define the training arguments
training_args = TrainingArguments(
    output_dir='./results',          # Output directory for model predictions and checkpoints
    num_train_epochs=3,              # Number of training epochs
    per_device_train_batch_size=128,  # Batch size for training
    per_device_eval_batch_size=128,   # Batch size for evaluation
    warmup_steps=500,                # Number of warmup steps for learning rate scheduler
    weight_decay=0.01,               # Strength of weight decay
    logging_dir='./logs',            # Directory for storing logs
    logging_steps=10,                # Log saving interval
    evaluation_strategy="epoch",     # Evaluate at the end of each epoch
    save_strategy="epoch",           # Save the model at the end of each epoch
    load_best_model_at_end=True,     # Load the best model when finished training (based on evaluation)
)
```

```python
# Define a simple function to compute metrics (e.g., accuracy)

def compute_metrics(eval_pred):
    logits, labels = eval_pred
    # Convert logits to a tensor if it's a numpy array
    if isinstance(logits, np.ndarray):
        logits = torch.tensor(logits)

    # Similarly, ensure labels are a tensor; this is not in your error but might cause issues later
    if isinstance(labels, np.ndarray):
        labels = torch.tensor(labels)

    predictions = torch.argmax(logits, dim=-1)  # Using 'dim' as the argument key for clarity
    accuracy = (predictions == labels).float().mean()
    return {'accuracy': accuracy.item()}

# Initialize the Trainer with the custom data collator

trainer = Trainer(
    model=model,
    args=training_args,
    train_dataset=train_dataset,
    eval_dataset=test_dataset,
    compute_metrics=compute_metrics,
    data_collator=custom_data_collator,  # Use your custom data collator
)

# Start training
trainer.train()

# Evaluate the model
results = trainer.evaluate()

# Save the model
trainer.save_model("./swin_v2_b_model_protest")

# You can also get the model's predictions on the test dataset using the following code:
# predictions = trainer.predict(test_dataset)

import os
import pandas as pd
import torch
from torchvision.transforms import functional as F
from transformers import Swinv2ForImageClassification, AutoImageProcessor
from concurrent.futures import ThreadPoolExecutor
from PIL import Image
from tqdm import tqdm

# Load the fine-tuned model and processor
model_dir = "./swin_v2_b_model_protest"
model_name = "microsoft/swinv2-base-patch4-window8-256"
model = Swinv2ForImageClassification.from_pretrained(model_dir)
processor = AutoImageProcessor.from_pretrained(model_name)

# Define a function to predict the class of a single image
def predict_image(image_info):
```

```python
    country_folder, image_file = image_info
    image_path = os.path.join(image_folder, country_folder, image_file)
    try:
        # Attempt to open the file as an image
        image = Image.open(image_path)

        # Ensure image is in RGB format
        if image.mode != 'RGB':
            image = image.convert('RGB')

        # Convert PIL image to PyTorch tensor
        image = F.to_tensor(image).float()
        inputs = processor(image, return_tensors="pt", do_rescale=False)
        outputs = model(**inputs)
        prediction = torch.argmax(outputs.logits, dim=-1)
        return {"Country": country_folder, "Image": image_file, "Prediction": prediction.item()}
    except Exception as e:
        print(f"Error processing file {image_path}: {str(e)}")
        return None

# Iterate through the files in the "images_by_country" folder and make predictions
image_folder = "images_by_country"
image_infos = []

for country_folder in os.listdir(image_folder):
    country_path = os.path.join(image_folder, country_folder)
    if os.path.isdir(country_path):
        for image_file in os.listdir(country_path):
            image_infos.append((country_folder, image_file))

# Use ThreadPoolExecutor to parallelize the image prediction
with ThreadPoolExecutor() as executor:
    predictions = list(tqdm(executor.map(predict_image, image_infos), total=len(image_infos), desc="Predicting"))

# Remove None entries (if any)
predictions = [pred for pred in predictions if pred is not None]

# Save the predictions to a CSV file
predictions_df = pd.DataFrame(predictions)
predictions_df.to_csv("predictions_by_country.csv", index=False)
```

## References

Bail, Christopher A. 2024. "Can Generative AI Improve Social Science?" *Proceedings of the National Academy of Sciences* 121(21):e2314021121. doi: 10.1073/pnas.2314021121.

Burke, Marshall, Anne Driscoll, David B. Lobell, and Stefano Ermon. 2021. "Using Satellite Imagery to Understand and Promote Sustainable Development." *Science* 371(6535):eabe8628. doi: 10.1126/science.abe8628.

Davidson, Thomas. 2024. "Start Generating: Harnessing Generative Artificial Intelligence for Sociological Research." *Socius* 10:23780231241259651. doi: 10.1177/23780231241259651.

Edelmann, Achim, Tom Wolff, Danielle Montagne, and Christopher A. Bail. 2020. "Computational Social Science and Sociology." *Annual Review of Sociology* 46(1):61–81. doi: 10.1146/annurev-soc-121919-054621.

Fan, Xinguang, and Yongjun Zhang. 2022. "'Just a Virus' or Politicized Virus? Global Media Reporting of China on COVID-19." *Chinese Sociological Review* 0(0):1–28. doi: 10.1080/21620555.2022.2116308.

Gilardi, Fabrizio, Meysam Alizadeh, and Maël Kubli. 2023. "ChatGPT Outperforms Crowd Workers for Text-Annotation Tasks." *Proceedings of the National Academy of Sciences* 120(30):e2305016120. doi: 10.1073/pnas.2305016120.

Jean, Neal, Marshall Burke, Michael Xie, W. Matthew Davis, David B. Lobell, and Stefano Ermon. 2016. "Combining Satellite Imagery." *Science* 353(6301):790–94. doi: 10.1126/science.aaf7894.

Lazer, David M. J., Alex Pentland, Duncan J. Watts, Sinan Aral, Susan Athey, Noshir Contractor, Deen Freelon, Sandra Gonzalez-Bailon, Gary King, Helen Margetts, Alondra Nelson, Matthew J. Salganik, Markus Strohmaier, Alessandro Vespignani, and Claudia Wagner. 2020. "Computational Social Science: Obstacles and Opportunities." *Science* 369(6507):1060–62. doi: 10.1126/science.aaz8170.

Lazer, David, Alex Pentland, Lada Adamic, Sinan Aral, Albert-László Barabási, Devon Brewer, Nicholas Christakis, Noshir Contractor, James Fowler, Myron Gutmann, Tony Jebara, Gary King, Michael Macy, Deb Roy, and Marshall Van Alstyne. 2009. "Computational Social Science." *Science* 323(5915):721–23. doi: 10.1126/science.1167742.

Ledford, Heidi. 2020. "How Facebook, Twitter and Other Data Troves Are Revolutionizing Social Science." *Nature* 582(7812):328–30. doi: 10.1038/d41586-020-01747-1.

Lin, Hao, and Yongjun Zhang. 2025. "The Risks of Using Large Language Models for Text Annotation in Social Science Research."

Liu, Ze, Yutong Lin, Yue Cao, Han Hu, Yixuan Wei, Zheng Zhang, Stephen Lin, and Baining Guo. 2021. "Swin Transformer: Hierarchical Vision Transformer Using Shifted Windows." *arXiv:2103.14030 [Cs]*.

Maher, Thomas V., Charles Seguin, and Yongjun Zhang. 2024. "The Racial Limits of Disruption: How Race and Tactics Influence Social Movement Organization Testimony before Congress, 1960–1995." *Social Forces* soae073. doi: 10.1093/sf/soae073.

Ollion, Étienne, Rubing Shen, Ana Macanovic, and Arnault Chatelain. 2024. "The Dangers of Using Proprietary LLMs for Research." *Nature Machine Intelligence* 6(1):4–5. doi: 10.1038/s42256-023-00783-6.

Seguin, Charles, Thomas V. Maher, and Yongjun Zhang. 2023. "A Seat at the Table: A New Data Set of Social Movement Organization Representation before Congress during the Twentieth Century." *Socius* 9:23780231221144598. doi: 10.1177/23780231221144598.

Wang, Boshi, Sewon Min, Xiang Deng, Jiaming Shen, You Wu, Luke Zettlemoyer, and Huan Sun. 2023. "Towards Understanding Chain-of-Thought Prompting: An Empirical Study of What Matters."

Wankmüller, Sandra. 2022. "Introduction to Neural Transfer Learning With Transformers for Social Science Text Analysis." *Sociological Methods & Research* 00491241221134527. doi: 10.1177/00491241221134527.

Won, Donghyeon, View Profile, Zachary C. Steinert-Threlkeld, View Profile, Jungseock Joo, and View Profile. 2017. "Protest Activity Detection and Perceived Violence Estimation from Social Media Images." *Proceedings of the 25th ACM International Conference on Multimedia* 786–94. doi: 10.1145/3123266.3123282.

Wu, Xiaogang, Yongjun Zhang, and Tianji Cai. 2024. *Computational Social Science: Application in China Studies*. Taylor & Francis.

Yao, Shunyu, Dian Yu, Jeffrey Zhao, Izhak Shafran, Thomas L. Griffiths, Yuan Cao, and Karthik Narasimhan. 2023. "Tree of Thoughts: Deliberate Problem Solving with Large Language Models."

Zhang, Yongjun, Hao Lin, Yi Wang, and Xinguang Fan. 2023. "Sinophobia Was Popular in Chinese Language Communities on Twitter during the Early COVID-19 Pandemic." *Humanities and Social Sciences Communications* 10(1):1–12. doi: 10.1057/s41599-023-01959-6.